# Non-Integer Dimension of Seasonal Land Surface Temperature (LST)

Sepideh Azizi [1], Tahmineh Azizi [2,*],

1 Department of Urban and Regional Planning, University of Illinois at Urbana-Champaign
2 Department of Mechanical Engineering, Florida State University

*Correspondence Email: tazizi@fsu.edu.

## Abstract

During few last years, climate change including global warming which is attributed to human activities and also its long-term adverse effects on the planet's functions have been identified as the most challenging discussion topics which have arisen many concerns and efforts to find the possible solutions. Since the warmth arising from Earth's landscapes affects the world's weather and climate patterns, we decided to study the changes in the Land Surface Temperature (LST) patterns in different seasons through non-linear methods. Here, we particularly want to estimate the non-integer dimension and fractal structure of the land surface temperature. For this study, the (LST) data has been obtained during the daytime by the Moderate Resolution Imaging Spectroradiometer (MODIS) on NASA's Terra satellite. Depending on what time of the year data has been collected, temperatures change in different ranges. Since equatorial regions remain warm, and Antarctica and Greenland remain cold, and also because altitude affects temperature, we selected Riley County in the U.S. state of Kansas, which does not belong to any of this type locations and we are interested to observe the seasonal changes in temperature in this county. The results of the present study show that the Land Surface Temperature (LST) belongs to the class of fractal process with non-integer dimension.

## Introduction

With increasing temperature around the world, photosynthesis and microbial and plant respiration and as a result releasing carbon dioxide to the atmosphere also increase [26]. During recent decades, climate changes and its long-term impacts such as accelerating ice melt in poles, warming oceans, and rising sea level and many other aspects of life entire earth has been widely recognized by many different organizations and researchers [5, 18]. According to the most recent Intergovernmental Panel on Climate Change (IPCC) report (IPCC, 2014 [38]) greenhouse gases (GHG) are at the highest level which caused the most significant changes in earth temperature in history [18]. Human activities have significantly raised the concentration of greenhouse gases in atmosphere and caused global warming which is affecting adversely all life of the earth [38, 44]. Climate change and global warming result in human activities by changing the Land Surface Temperature (LST) make significant changes in ecosystem structure and consequently function [43, 50]. For example, they cause acceleration in the environmental status changes on the Tibetan Plateau (Zhong et al. (2011)) [50]. Many studies have also documented the effect of land-use and land-cover change which are caused by human activities on land–atmosphere interactions (Boisier et al., 2013, Boisier et al., 2012, de Noblet-Ducoudr´e et al., 2012) [12, 13, 17].

States at Risk which is a project to demonstrated the effects of climate changes in all 50 states of United States, divides the climate changes impacts into five categories: extreme heat, drought, wildfires, coastal flooding and inland flooding. There are many real-world examples of these five impacts of climate changes in different parts of the earth. For example, the devastating brush-fires happened in Australia in 2019 because of high temperature is one the most dramatics recent disasters result in climate changes which followed by extreme heat and long drought beside killing people and animals (Source(s): SafeHome.org, States at Risk). Across the United States, we can see different impacts of climate changes in daily life such as extreme heat during summer and intolerable cold during winter, destructive storms, wildfires prolonged bush-fires in plains and many more that have a significant impact on economy and other aspects of lives.

The Land Surface Temperature (LST) plays a key role in environmental studies and climate researches [18]. The (LST) demonstrates the temperature of any surface of the Earth in a specific location. Since, a MODIS satellite collects the data through atmosphere to the ground, this surface can be snow, ice, grass, leaves of trees, roof of buildings. Therefore (LST) is not the same as the air temperature that we hear in the daily weather report. (LST) is a combination of vegetation and bare soil temperatures and because these two factors vary very fast in response to any solar radiation and or aerosol variations, (LST) demonstrates rapid changes as well. Scientists by collecting this data want to find the relationship between increasing atmospheric greenhouse gases and land surface temperature, and generally speaking, discovering the impact of increasing (LST) on glaciers, ice sheets, permafrost, and the vegetation in Earth's ecosystems (Source(s): Moderate Resolution Imaging Spectroradiometer (MODIS)).

In the United States, southwest including central plains in Kansas confront unprecedented drought. Increasing the earth temperatures and reduction in rainfall will cause more extreme drought in the future than before effects adversely the urban dwellers and also farmers across this state. Using the data published by Climate Central (an independent organization with this purpose to conduct research about the climate changes and its impact on the public), the state of Kansas has been faced with extreme heat, drought, wildfires, and inland flooding result in global warming (Source(s): Climate Central, SafeHome.org, States at Risk). In figure (1), we can see the seasonal changes in Land Surface Temperature (LST) of Riley County in the U.S. state of Kansas obtained via Google Earth Engine, from winter 2021-winter 2022.

Climate is a complex system since not all the variables in this system can be observed and because of uncertainty, our knowledge to predict is limited. To possibly describe complex natural processes, Benoit Mandelbrot in 1983 introduced fractals as immensely useful object in numerous scientific fields [34]. Mathematically speaking, fractals are infinitely irregular self similar objects which are characterized by power law or scaling exponent over a wide range of scales [33, 34]. Thanks to Mandelbrot' research in fractal geometry, earth scientists are able to make probability and predict the size, location, and time of future natural disasters by measuring past events. If a certain process can be characterized by a single scaling exponent, then it is called mono-fractal. However, if a phenomena requires a large number of scaling exponents (can be infinite) to characterize their scaling structure, then it is called multi-fractal [27, 31, 42]. In studying real world time series data, we may deal with databases which exhibits power law or self similarity a wide range of size and time scales. For this type phenomena, we need to apply non-linear techniques to characterize the complexity and self similarity of data because the traditional time series techniques fail to classify databases with a variety range of scaling features [5–8, 30, 35].

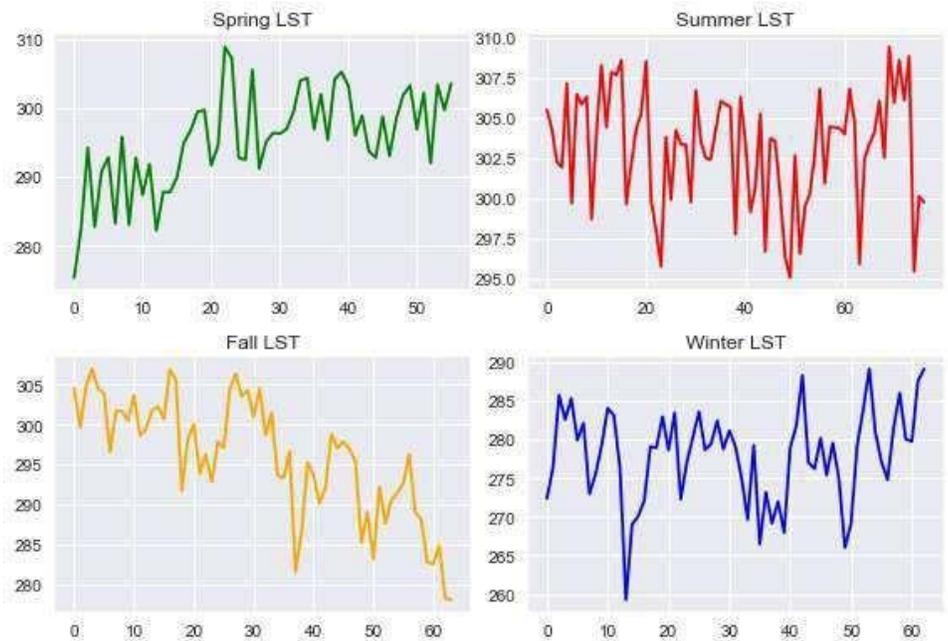

Figure 1. Seasonal trends in Land Surface Temperature (LST) of Riley County in the U.S. state of Kansas obtained via Google Earth Engine, started from winter 2021-winter 2022.

The present study investigates the self-similarity and power-law structures of the Land Surface Temperature (LST) time series data using some non-linear techniques and fractal geometry. We start with statistical regression analysis on the Land Surface Temperature (LST) databases of a selected region, the Riley County in the U.S. state of Kansas. This database has been taken using Google Earth Engine, the Moderate Resolution Imaging Spectroradiometer (MODIS) on NASA's Terra satellite. Because of the irregularity in the nature of the (LST) database in different days of the year, we use the non-linear methods such as power-law analysis, fractal dimension analysis and multifractal spectrum technique to characterize the structure of data. The vibration analysis using power spectral densities (PSD) method has been performed to find whether some type of power-law scaling exists for various statistical moments at different scales of these databases. Then, we use Discrete Wavelet Transform (DWT) and Wavelet Leader Multifractal (WLM) analysis to discover the possibility that these time series databases belong to the class of multifractal process for which a large number of scaling exponents are required to characterize their scaling structures. Finally, a non-linear analysis called the Fractal Dimension (FD) analysis using Higuchi algorithm has been carried out to quantify the fractal complexity of data.

## Materials, Methods and results

### Study Area

The study area is Riley County which is located in the U.S. state of Kansas. According to the U.S. Census Bureau, its total area is 622 square miles (1610$km^2$), including 610 square miles (1600 $km^2$) land and 12 square miles (31 $km^2$) (2.0%) water (Source(s): US Gazetteer files: 2010, 2000, and 1990. United States Census Bureau. February 12, 2011.

Retrieved April 23, 2011). Figure 2-(a) displays the solar radiation map of Riley County (it describes how much this area receives sunlight or radiation) which is obtained via online geospatial tools. In figure 2-(b), we can see the Digital Elevation Model (DEM) of Riley County which represents the bare ground topographic surface of the county but it is excluding trees, buildings, and any other surface objects.This map is created via USGS DEM which is primarily obtained from topographic maps. Part (c) shows how we derived (LST) values for each grid cell from a 16-day time-span during January 2021 to January 2022. The grids are produced using Aqua Level-3 Moderate Resolution Imaging Spectroradiometer (MODIS) Version 5 global daytime and nighttime (LST) 16-day composite data (MYD11A2). Part (d) of figure 2, exhibits a 3D model of the Digital Elevation Model (DEM) of Riley County derived from topographic maps.

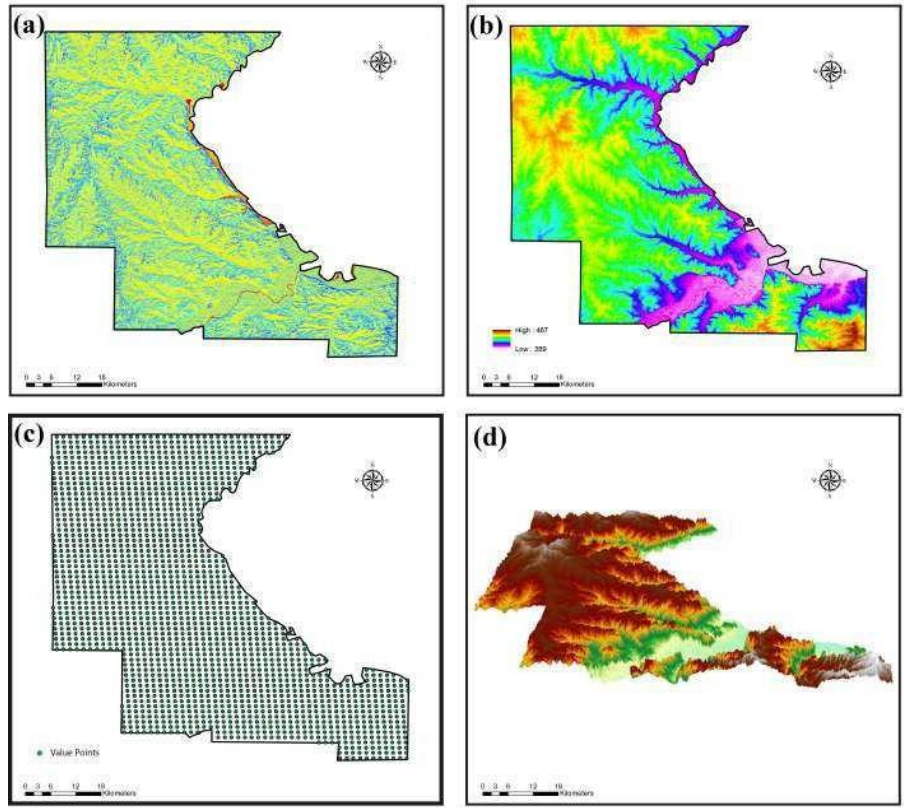

Figure 2. Study area of the Riley County (a) Solar radiation map obtained via online geospatial tools; (b) shows the Digital Elevation Model (DEM) of Riley County derived from topographic maps; (c) (LST) grid cells; (d) 3D model of the Digital Elevation Model (DEM) obtained via topographic maps.

DEMs are the most common cell-based representations tools in studying the shape of the earth's surface which are used extensively to characterize the land surface features (Source(s): The sole science agency for the Department of the Interior (USGS)).

## Data: The Land Surface Temperature (LST) of Riley County for February, April, July, and October (2021)

The present study investigates the non-linear patterns of daytime (LST) databases which have been collected using Google Earth Engine and over Riley County in the U.S. state of Kansas. Google Earth Engine is a combination of multi-petabyte catalog of satellite imagery and Geo-spatial databases with planetary-scale analysis capabilities [20]. Google Earth Engine helps to track changes and or measuring the differences on the Earth's surface and has been used widely to find map trends. (LST) databases are remotely sensed data and taken from MODerate Resolution Imaging Spectroradiometer (MODIS) sensor in the form of 16-day composites of daytime values at a resolution of one km during February, April, July and October (2021) . See figures (3)-(17).

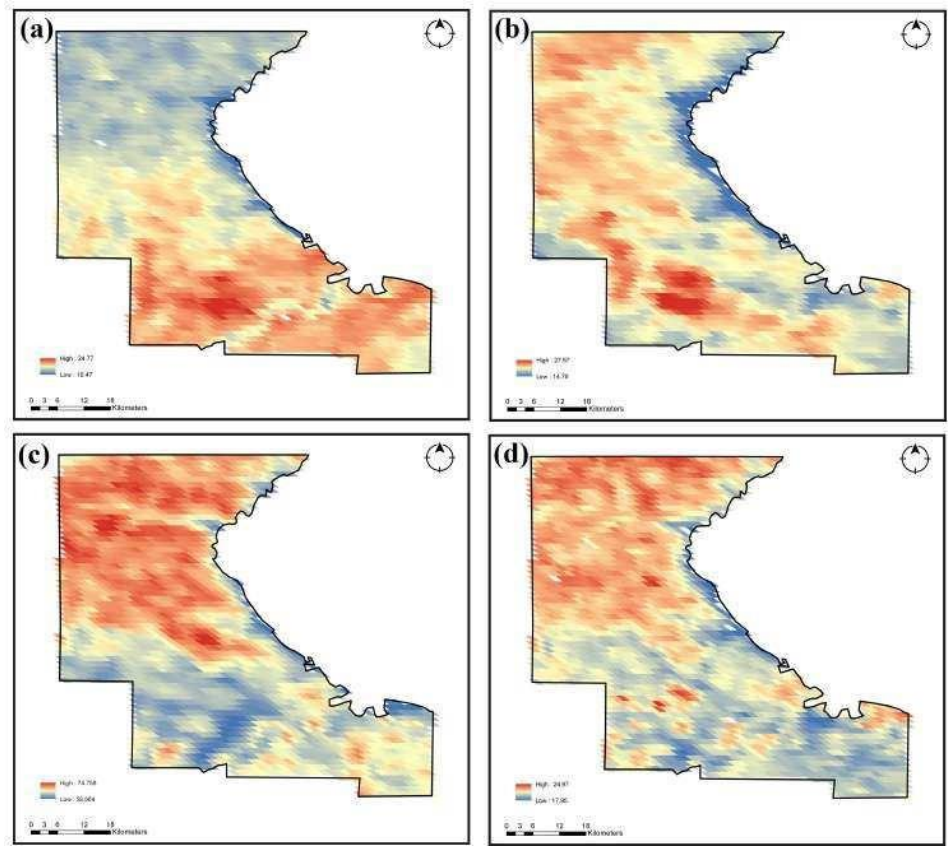

Figure 3. Study area of the Riley County and Land Surface Temperature (LST) maps (a) The (LST) map of February 2021; (b) The (LST) map of April 2021; (c) The (LST) map of July 2021; (d) The (LST) map of October 2021.

Time series regression analysis

To explore and quantify the trends and patterns in (LST) database, we perform a time series regression analysis. Time series regression is a statistical technique which uses experimental data to predict the future (LST) based on the current (LST) responses by transferring the dynamics of our predictors which is the height. This forecasting method provides a great understanding of the structure of the dynamic system which is (LST) database [21].

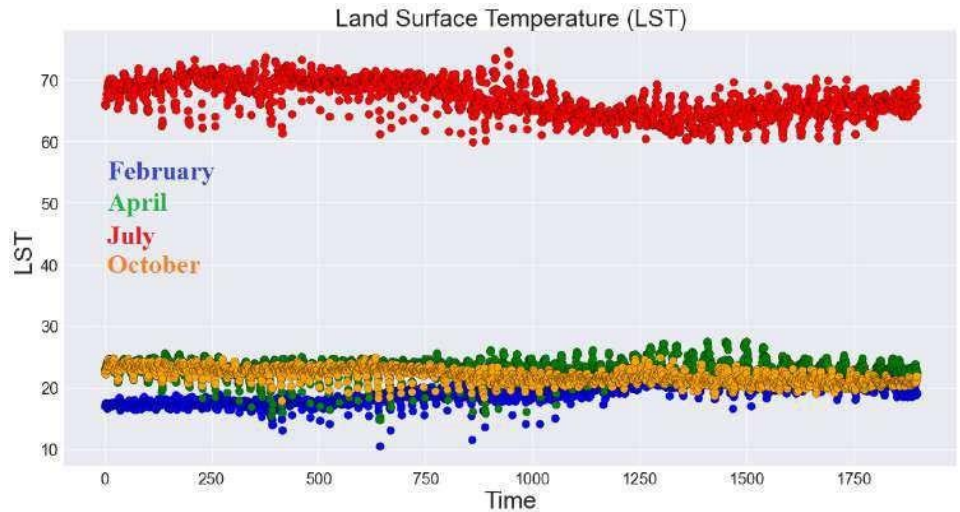

Figure 4. The scatter plot of Land Surface Temperature (LST) of Riley County during February, April, July and October (2021) obtained via Google Earth Engine.

We start with defining the design matrix $H_t$ of current and past height data and ordered bt time and we call this matrix as regressor matrix. Next, we use ordinary least squares (OLS) method for the linear regression (LR) equation:

$$Lt = Ht\,\beta + \epsilon t \qquad (1)$$

where $\beta$ is the linear estimator parameter, $L_t$ represent the (LST) database. The model (1) helps to find the linearity between the (LST) variable $L_t$ and regressor matrix $H_t$.

If we replace the linear part of equation (1) with a non-linear function $f(H_t, \epsilon_t)$, we can better understand the relationship between the (LST) variable $L_t$ and regressor matrix $H_t$ as the following form:

$$Lt = f(Ht, \epsilon t) \qquad (2)$$

See figure (2).

Time-Frequency Analysis using Continuous Wavelet Transform (CWT) and Wavelet-Based Semblance Analysis

When we are working with stationary databases with constant frequencies throughout the time intervals of data, Fourier transform helps to visualize the data over time, however, this technique fails when we have datasets which are changing in time in response to a particular stimulus, i.e. they have time varying features [15, 22]. The reason is the main limitation of Fourier transform which gives a broad power spectrum by integrating the frequency of time series data along the whole data. Because our Days (LST) databases are non-stationary, we will apply time-frequency methods such as Continuous Wavelet Transform (CWT). Wavelet-based approaches provide the ability to account for temporal (or spatial) variability in spectral character. Although wavelet analysis is relatively new compared with Fourier analysis, its use has become widespread in recent years, and so the theory will be described only briefly here [16]. Mallat (1998) or Strang and Nguyen (1996) contain excellent and detailed summaries of wavelet analysis. The continuous wavelet transform (CWT) of a dataset h(t) is given by (Mallat,

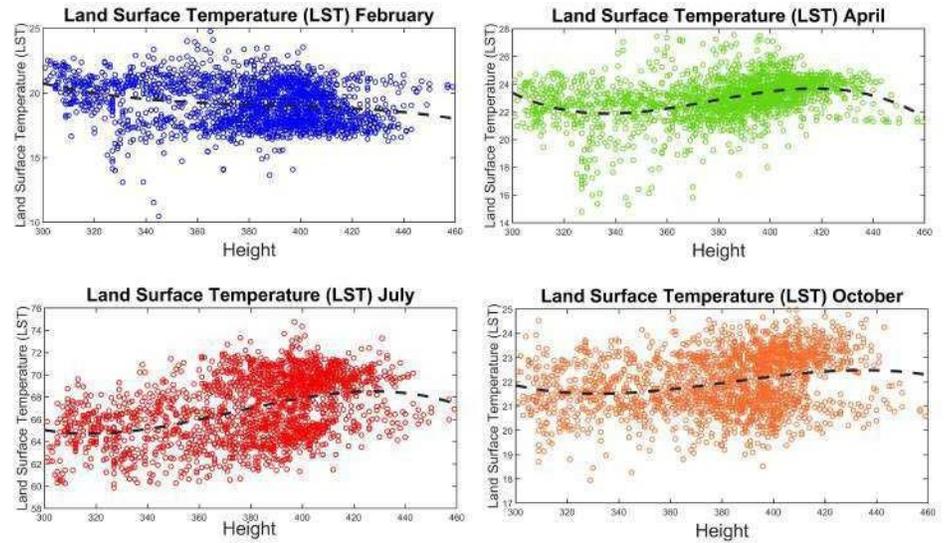

Figure 5. Time series non-linear regression model (2) using the Land Surface Temperature (LST) of Riley County during February, April, July and October (2021) obtained via Google Earth Engine.

1998) [16, 32] $CWT(u,s)$  (3)

where $s$ is scale, $u$ is displacement, $\Phi$ is the mother wavelet used, and $*$ means complex conjugate. The CWT is therefore a convolution of the data with scaled version of the mother wavelet. Of course, the time coordinate $t$ in equation (3) could equally well be the spatial coordinate $x$ if profile data were being analyzed.

Similarity measures are becoming increasingly commonly used in comparison of multiple datasets from various sources. Semblance filtering compares two datasets on the basis of their phase, as a function of frequency. Semblance analysis based on the Fourier transform $ET(f)$ [11] where $h(t)$ is dataset, $f$ represents the frequency and $t$ is the time, suffers from problems associated with that transform, in particular its assumption that the frequency content of the data must not change with time (for time-series data) or location (for data measured as a function of position). To overcome these problems, semblance is calculated here using the continuous wavelet transform. When calculated in this way, semblance analysis allows the local phase relationships between the two datasets to be studied as a function of both scale (or wavelength) and time [16, 45].

When we define the Fourier transform (4) of time series datasets $h_1(t)$ and $h_2(t)$, the semblance $S$ or the difference in the phase angle at each frequency can be obtained using the following formula $s = cos\theta(f)$ [14, 47] where $R_1(f)$ and $R_2(f)$ represent the real parts of the Fourier transform of dataset 1 and dataset 2 respectively, $I_1(f)$ and $I_2(f)$ demonstrate the imaginary parts of the

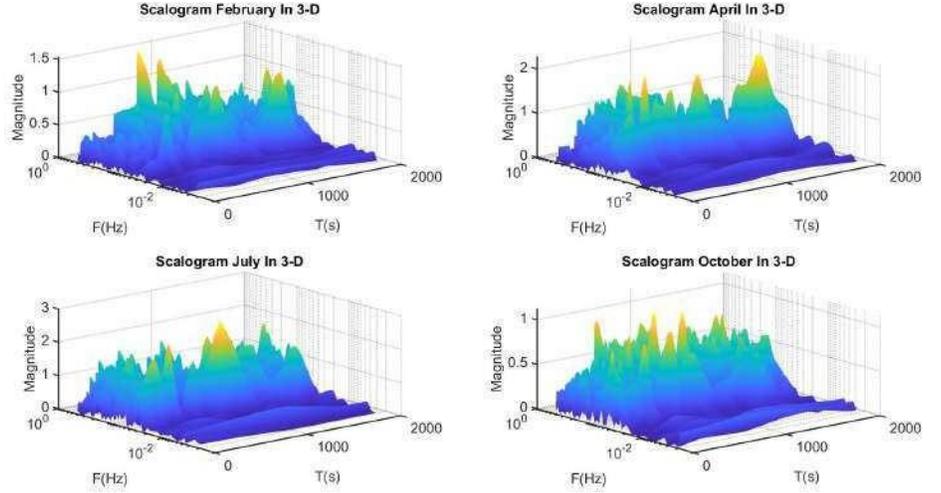

Figure 6. Time-frequency representations of the Land Surface Temperature (LST) of Riley County during February, April, July and October (2021) using Continuous Wavelet Transform (CWT) in three dimensional Time-Frequency-Magnitude space.

Fourier transform of dataset 1 and dataset 2 respectively, and the semblance *S* varies between −1 to +1. If *S* = +1, there exist a perfect phase correlation, for *S* = 0 there is no correlation, and for *S* = −1 we have a perfect anti-correlation. However, the the Fourier-based semblance method (5) does not work well in real world application fails to present the correlation between time series data. Therefore, for our study to find the correlation between two time series data, we use a wavelet based method called the cross-wavelet transform which is introduced by (Torrence et al. (1998)) has the form [45]:

$$CWT_{a,b} = CWT_a \times CWT_b^* \qquad (6)$$

This is a complex quantity and its amplitude which is called the cross-wavelet power defined as

$$AMP = |CWT_{a,b}| \qquad (7)$$

with local phase $\theta$

$$\arctan \frac{Im(CWT_{a,b})}{Re(CWT_{a,b})}$$

The local phase $\theta \in (-\pi, \pi)$ measures the phase correlation between two time series data. Now, we define the wavelet-based semblance as the form

$$S = \cos^n(\theta) \qquad (8)$$

Here *n* > 0 is odd and integer. For *S* = −1, our time series datasets are inversely correlated, for *S* = 0 there are un-correlated, and for *S* = +1, they are correlated. However, the wavelet-based semblance (8) fails to demonstrate the correlation between two noisy time series data because of of the lack of amplitude information [16]. Therefore, we define the vector dot product of the two complex wavelet vectors at each point in scale and position [16]

$$D = \cos^n \theta \, |CWT_a \times CWT_b^*| \qquad (9)$$

As we can see from equation (9), *D* is a combination of the wavelet-based semblance (8) and the cross-wavelet power (7), which helps to find the phase correlations of the larger amplitude components of the noisy dataset.

In figure (7), we have plotted the Dot product *D* of Land Surface Temperature (LST) of Riley County (figure (4)) during February, April, July and October (2021) using the formula (9) where dark red corresponds to large-amplitude signals with semblance equals +1, and yellow corresponds to large-amplitude signals with positive semblance between 0 and +1. The semblance results show that (LST) dataset and height datasets are highly correlated.

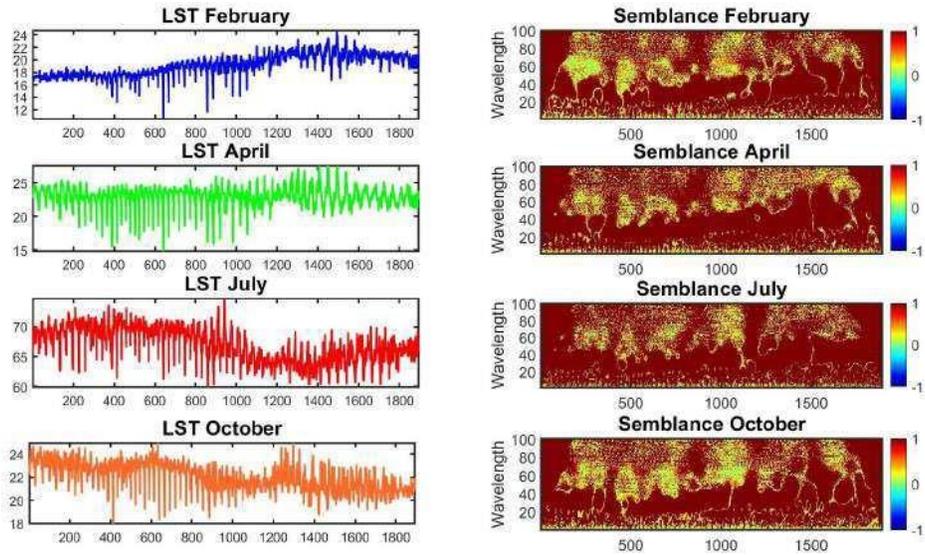

Figure 7. Semblance of (LST) datasets shown in figure (4) to find the correlation between (LST) data and height of each grid point in figure (2) computed using the vector dot product of the two complex wavelet vectors (9), the correlation varies as a function of time and wavelength. Strongly correlated on larger wavelengths ($S = +1$) in dark red, Positively correlated ($0 < S < +1$) in yellow, zero correlation ($S = 0$) in green. The results show mostly dark red, implying strong positive correlation between (LST) data and height of each grid point in figure (2).

Vibration Frequency Analysis Via Power Spectral Densities (PSD)

Power spectral density (PSD) analysis is one of the methods in frequency-domain analysis which has been used frequently to analyze the signal vibration. Power defines the magnitude of the PSD which is the mean square value of the given time series data. PSD as a function of frequency demonstrates the distribution of a time series data over a spectrum of frequencies and its magnitude is normalized to a single Hertz bandwidth. PSD curves display the continuous probability density function of each measure, and it is obtained via multiplying each frequency bin of fast Fourier transform (FFT) by its complex conjugate to get a real spectrum. Because, the nature of (LST) database is non-linear and non-stationary, the welch (PSD) method with overlapped segmentation which is an averaging estimator technique has been applied to study the complex fluctuations in (LST) structures. Then, we use least square approximation technique to fit a linear regression model to the logarithm of power spectral density of (LST) database. Finally, we compute the slope for each regression model of (LST) data. In figure (8), we have shown the fitted least squares approximation to the logarithm of power spectral density of the Land Surface Temperature (LST) of Riley County during February, April, July and October (2021).

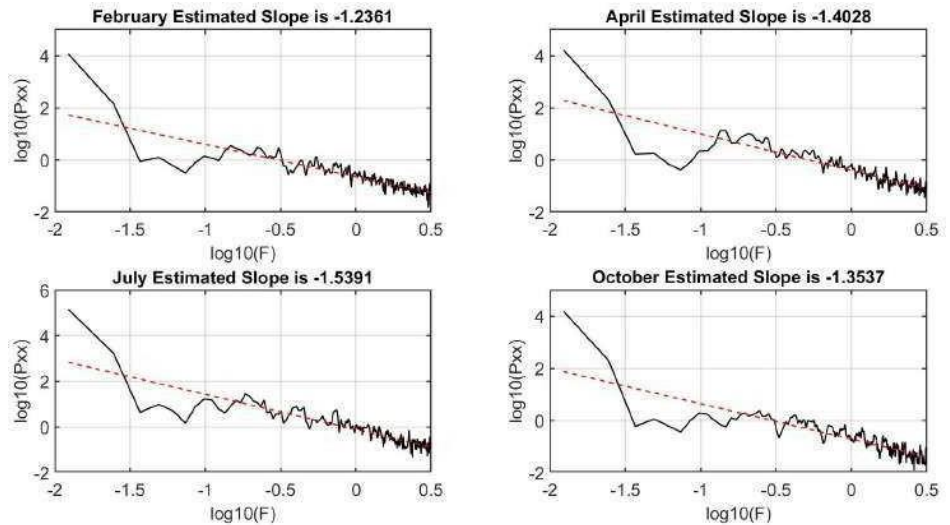

Figure 8. Power-Law or self-similarity patterns; Fitted least squares approximation to the logarithm of power spectral density of the Land Surface Temperature (LST) of Riley County during February, April, July and October (2021) obtained by wavelet techniques.

In fractal processes, we can find a scaling relationship between power and frequency $f$ in the spectral domain. The PSD results in figure (8) represent the fractal structure in (LST) database via a linear, negative slope of fitted least square lines. As result, the (LST) time series data cannot be obtained by one or a finite set of subsystems, and different components in each fractal process act at different time scales. Even though, PSD revealed the power law structure in (LST) databases, however, it failed to classify these four groups of data.

### Fractal Geometry and World's Weather and Climate Patterns

Multi-fractal Analysis and Discrete Wavelet Transform (DWT)

We have shown that complex structures in (LST) databases in figure (4) exhibit long-range correlation, i.e. they are self-similar: statistically, vibrations at one time scale are similar to those at multiple other time scales. To discover whether these four groups of (LST) data belong to the class of multi-fractal processes or they are mono-fractal, we have plotted the scaling exponents of the Land Surface Temperature (LST) of Riley County during February, April, July and October (2021) in figure (9).

The non linear relationships between scaling exponents and the statistical moments could be a sign of multi-fractality in the structure of data, however, we need to check this claim via multi-fractal analysis. Multi-fractal analysis helps us to understand weather the (LST) databases are multi-fractal, i.e. means that a larger number of scaling exponents are required to characterize the scaling structures of time series data. To investigate scaling law of a multi-fractal process, the standard partition multi-fractal function was first introduced which was based on generalized fluctuation analysis. However, it failed to fully characterize the scaling behavior of the non-stationary time series data [10]. Arneodo et al., 1995b [4], Bacry et al., 1993 [9], Muzy et al., 1993, 1994 [36], developed a statistical techniques based on the Continuous Wavelet Transform (CWT) to characterize the multi-fractal description of singular measures. Arneodo et al.,

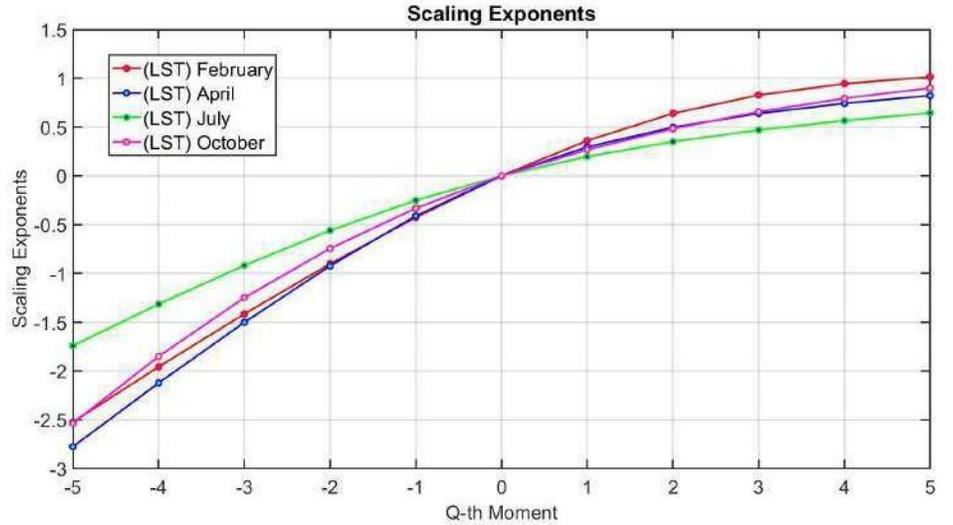

Figure 9. Scaling exponent of power spectral density of the Land Surface Temperature (LST) of Riley County during February, April, July and October (2021).

2002 developed a method called Wavelet Transform Modulus Maxima (WTMM) which solved the deficiency of partition multi-fractal function and provided a wide insight into a variety of problems. Abry et al., 2000, 2002a,b [1–3], Veitch and Abry, 1999 [46], developed multi-fractal analysis based on Discrete Wavelet Transform (DWT) including the recent use of Wavelet Leaders (WL) (Jaffard et al., 2006 [28], Wendt and Abry, 2007 [48], Wendt et al., 2007 [49]). In this study, we focus on multi-fractal analysis using (DWT) and (WL) [23, 24, 28, 40, 41, 48, 49]. The wavelet analysis associates the dimension of the fractal sets to Hölder exponent $H(\tau)$ to quantify the spectrum of singularity of the pointwise regular function $F$ [28]. The Hölder exponent of a fractal process $F(\tau)$ can be defined as follows:

Definition 0.1. *[41] A fractal process $F(\tau)$ satisfies a Hölder condition, when there exist $H(\tau) > 0$, such that*

$$|F(\tau_1) - F(\tau_2)| \simeq |\tau_1 - \tau_2|^{H(\tau)} \tag{10}$$

*We can find $H(\tau)$ for constant F from the coarse Hölder exponents as*

$$h_\xi(\tau) = \frac{1}{\log \xi} \log \sup_{|\tau_1 - \tau_2| < \xi} |F(\tau_1) - F(\tau_2)| \tag{11}$$

The following sets may be defined to extract the geometry of a signal

$$E_h^{[d]} = \{\tau : H(\tau) = d\} \tag{12}$$

with varying *d*, these sets describe the local regularity of signal. We call the map

$$d \mapsto \dim(E^{[d]}) \tag{13}$$

which is a compact form of the singularity structure of the fractal process $F$, the multi-fractal spectrum of $F$ [41]. In a global setting, to describe the complexity of a signal, we may need to count the intervals over which the fractal process $F$ evolves with Hölder exponent $H(\tau)$ and it gives an estimation of $\dim(E^{[d]})$. The grain exponent which is a discrete approximation to $h_\xi(\tau)$ can be written as the following form [41]

$$h_k^{(n)} := -\frac{1}{n} \log_2 \sup\{|F(\eta) - F(\tau)| : (k-1)2^{-n} \leq \eta \leq \tau \leq (k+2)2^{-n}\} \tag{14}$$

Therefore, the grain multi-fractal spectrum has the form [23, 24, 40]

$$F(d) = \lim_{\xi \to 0} \lim_{n \to \infty} \frac{\log N^n(d, \xi)}{n \log 2} \quad (15)$$

where

$$N^n(d, \xi) = \#\{k : |h_k^{(n)} - d| < \xi\} \quad (16)$$

From multi-fractal analysis in figure (10), we can easily see a clear loss of multi-fractality in all four groups of (LST) time series data, i.e. they are homogeneous and mono-fractal since their spectrum displays a narrow width of scaling exponent.

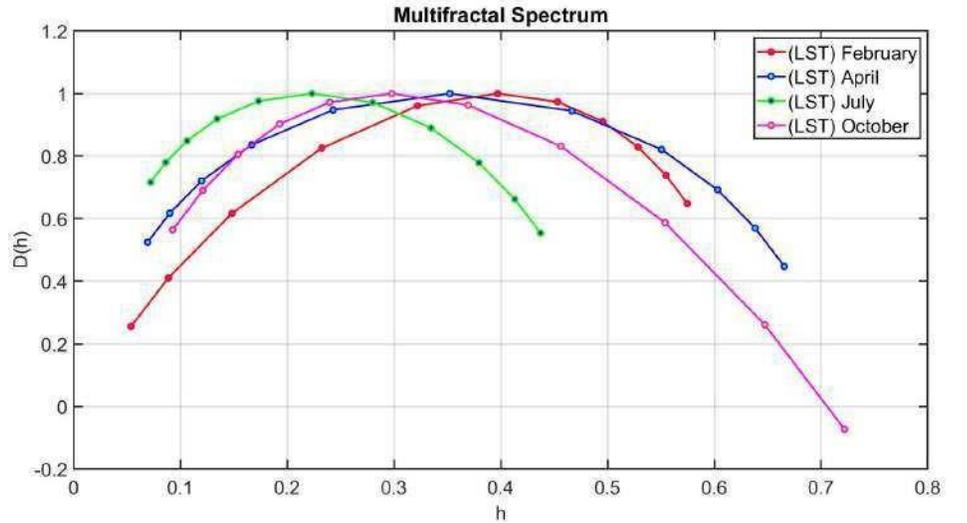

Figure 10. The multi-fractal spectrum analysis of the Land Surface Temperature (LST) of Riley County during February, April, July and October (2021) shows occurrence of mono-fractal process with a narrow range of exponents for all four groups of (LST) time series data.

Higuchi Fractal Dimension Algorithm

One of the key concepts in theory of fractal geometry is the concept of fractal dimension. There are a variety of techniques to approximate the dimension of an irregular object. For example, the dimension of an object can be obtained via covering it with small boxes. The Hausdorff-Besicovitch dimension or box-counting dimension is one of the most often used methods to estimate the fractal dimension which works in this way [39]. To find box-counting dimension, we consider a square $L \times L$ and partition it into grids of linear length $\epsilon$. If $N = (L/\epsilon)^2$ defines the total number of boxes, then, the total measure would be [19]:

$$M = M_\mu = L^2 \epsilon^{\alpha - 2} \quad (17)$$

where we denote the measuring unit by $\mu = \epsilon^\alpha$ and try out a number of different $\alpha$. If we choose $\alpha = 1$ then $M \to \infty$ as $\epsilon \to 0$. This means that the length of a square is infinite, which makes sense. If we try $\alpha = 3$, then $M \to 0$ as $\epsilon \to 0$. This means that the volume of a square is 0, which is also correct. For $\alpha = 2$, we can find the true dimension of square. Using this argument, we define the Hausdorff-Besicovitch dimension as follows: first, we need $\alpha$-covering measure, which is the summation of the total measure in all the "boxes" denoted by $V_i$, $i = 1, 2, \ldots$, subject to the condition that the union of the boxes covers the object $E$, while the size of the largest $V_i$ is not greater than $\epsilon$..

Although, this method can successfully measure the self-similarity of a fractal process, however, it fails when the sudden changes happen in the irregular time series data sets [37]. To solve this problem, a variety of different non-linear methods such as Higuchi algorithm, power spectrum analysis, and Katz algorithm have been developed by different researchers [25, 29]. To find the fractal dimension of each (LST) timeseries data, we apply the Higuchi Algorithm [25]. At first, we define a finite time series

$$X_1, X_2, X_3, \ldots, X_N$$

and we create $k$ new time series $X^k_m$ of the form

$$X_m, X_{m+k}, X_{m+2k}, \ldots, X_{[m+A\,k]}$$

such that $A = (N - m)/k$. For each time interval $k$, and the initial time $m = 1, 2, \ldots, k$, we compute the length of $X^k_m$

here, $R = (N - 1)/[A]k$ called the curve length normalization factor. Next, we calculate the mean of $L^k_m$ for $m = 1, 2, \ldots, k$ to find the average of curve length for each $k = 1, \ldots, k_{max}$. Then, we plot $\log(L^k_m)$ versus $\log(1/k)$ for different time interval $k$. Finally, we use the least-squares approximation to estimate the slope of each regressed line as Higuchi fractal dimension for an optimal value of time interval $k = 500$ (optimal means there is no change in Higuchi fractal dimension after this value).

We have estimated the fractal dimension of the Land Surface Temperature (LST) of Riley County during February, April, July and October (2021) and plotted their regression models for each time series in figure (11).

From the results in figure (11), we can compare the fractal dimension of different (LST) data as follows

$$FD_{\text{February}} < FD_{\text{July}} < FD_{\text{October}} < FD_{\text{April}}$$

Although, fractal dimension using the Higuchi algorithm is a good index for comparing the self-similarity and complexity in fractal structure of different (LST) data sets, however, it fails to classify four groups of (LST) data sets.

## Discussion

During recent decades, the rapid modification in land's surface by human activities has caused numerous environmental impacts. Land surface plays an important role in the structure of climate system and many related organizations put so much efforts to build observational land surface systems. Since controlling the Land Surface Temperature (LST), strongly impact the climate change in local and global scale, we decided to analyze the complexity of (LST) database to better understanding its pattern and structure. In the present study, we used satellite data of the Moderate Resolution Imaging Spectroradiometer (MODIS) of the Land Surface Temperature (LST) of Riley County during February, April, July and October (2021) and we conducted our research within the framework of fractal geometry.

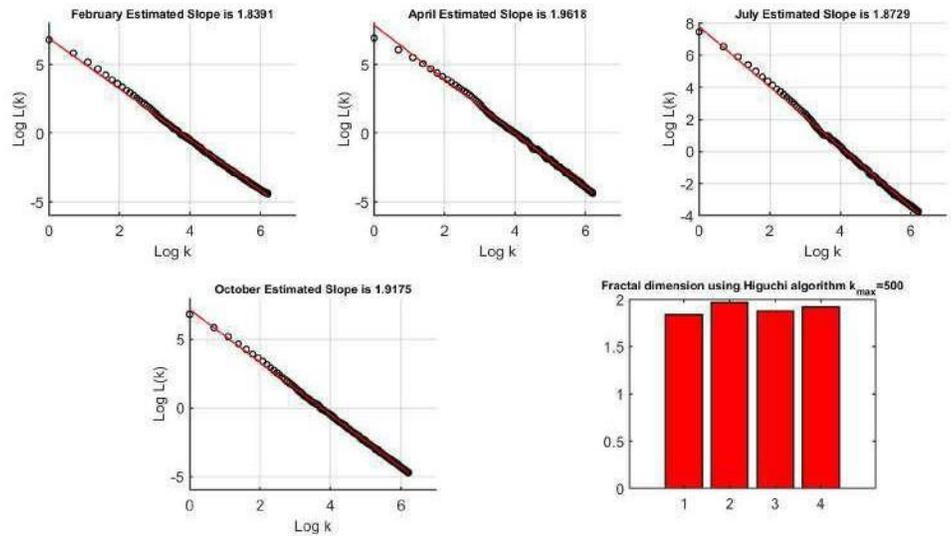

Figure 11. Plots of log($L_m^k$) versus log($k$) for time interval $k = 500$, the logarithmic scale and the corresponding slope of fitted regression line (the Higuchi fractal dimension) of the Land Surface Temperature (LST) of Riley County during February, April, July and October (2021).

To explore the differences in the fractal patterns and complex dynamics of different (LST) data sets, we established our analysis based on different quantitative and analytical non-linear methods to find a computational framework in classification of (LST) databases. We started with vibration analysis using power spectral densities (PSD) method to estimate the exponent from realizations of these processes and to find out if the data of interest exhibits a power-law PSD. We performed multifractal analysis using discrete wavelet methods to discover whether some type of power-law scaling exists for various statistical moments at different scales of (LST) databases and also to explore if our database follows multifractal structure for which a large number of scaling exponents are required to characterize their scaling structures. Finally, we computed the Higuchi fractal dimension of (LST) databases for each group to compare the complexity of (LST) databases in different time intervals.

This study revealed that the (LST) databases are mono-fractal, i.e. their spectrum displays a narrow width of scaling exponent, or in another word, a single global exponent is enough to characterize the fractal behavior and scaling properties of time series data. This important and novel finding helps to better understanding the structure of (LST) databases and consequently, better controlling the climate change caused by human activities. Although, because of limited number of databases, the vibration analysis using (PSD), fractal dimension and multi-fractal analysis could not classify (LST) databases for different seasons, however, they may be used as a comprehensive framework to further analysis, characterize and compare the fractal behavior of (LST) time series and extracting their complex structure. A future direction for this study would be utilizing larger number of databases from different climate, time scales and location to support our conclusion.

## Appendix (Traditional Statistical Analysis)

In this study, we applied the time series linear regression to find the linear relationship between the (LST) variable $L_t$ and regressor matrix $H_t$ in model (1) using the Land Surface Temperature (LST) data in figure (4). In figure (12), we can see the time series linear regression analysis of the seasonal (LST) of Riley County during February, April, July and October (2021) obtained via Google Earth Engine in figure (4). Here, we analyze the time series linear regression in figure (12) to test if the linear model works well for the Land Surface Temperature (LST) data or not. At first, we check that if residuals have non-linear patterns or not. We have plotted the Residuals vs Fitted graph for February, April, July and October (2021) (4) in figures (13)-(16) separately. According to the Residuals vs Fitted results, we have almost equally spreading residuals around the horizontal line (no distinct patterns), that means there is no non-linear relationships in any of time series regression models.

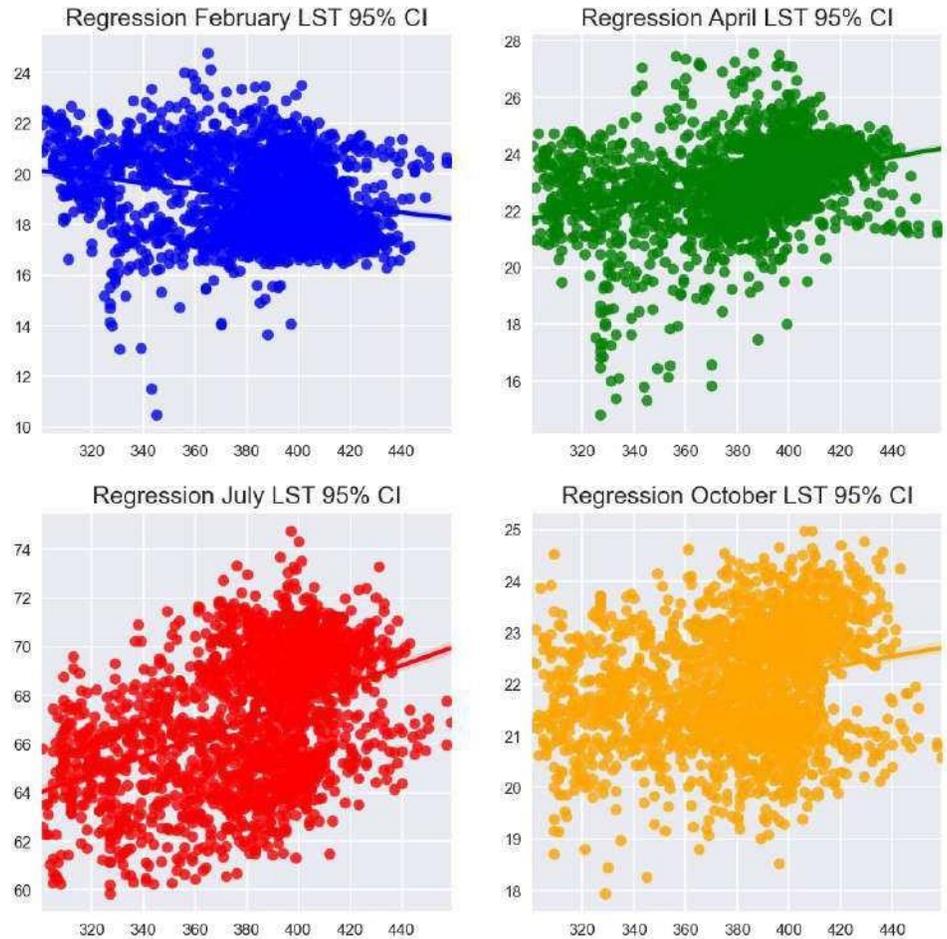

Figure 12. Time series linear regression model (1) using the Land Surface Temperature (LST) of Riley County during February, April, July and October (2021) obtained via Google Earth Engine (see figure (4)).

Secondly, we are interested to check if residuals are normally distributed. Therefore, we plot Normal quantile-quantile (Q-Q) graphs of Land Surface Temperature (LST) data (4) in figures (13)-(16) separately. In fact, Q-Q plots are scatter plots which are produced when we plot a set quantile against another one. As we can from the Q-Q plots, residuals do not follow a straight line very well and they deviate. As a result, the Land Surface Temperature (LST) data came from a Normal theoretical distribution or (LST) data sets are not Normally distributed (we can see bi-normality in figure (17)).

Next, we plot Spread-Location plots of the Land Surface Temperature (LST) data (4) to check if we have equal variance or homoscedasticity in our database, i.e. if residuals are spreading equally or not. From figures (13)-(16), the residuals are almost appeared to be randomly spread (do not spread widely).

Finally, we plot the Residuals vs Leverage graphs of the Land Surface Temperature (LST) data (4) in figures (13)-(16) to find influential subjects to determine the time series regression model. As we can easily see from the Residuals vs Leverage graphs, there is no obvious Cooks distance lines since all subjects are well located inside of the Cook's distance lines. Thus, we do not have any influential observation in (LST) databases.

The diagnostic plots showed that the residuals of the linear model (1) are not normally distributed, therefore, the time series linear regression model may be unreliable to represent the Land Surface Temperature (LST) data (4). It is true that there exist many other different techniques such as $R^2$, slope coefficients, and p-values to test the regression results, but, they may not be able to check all aspects of a perfect model. However, the residuals can be considered as indicator to characterize if a model works perfectly or not. Moreover, residuals can well represent the unrevealed patterns in the data and need to be considered as a good platform to characterize all possible dynamics in different databases.

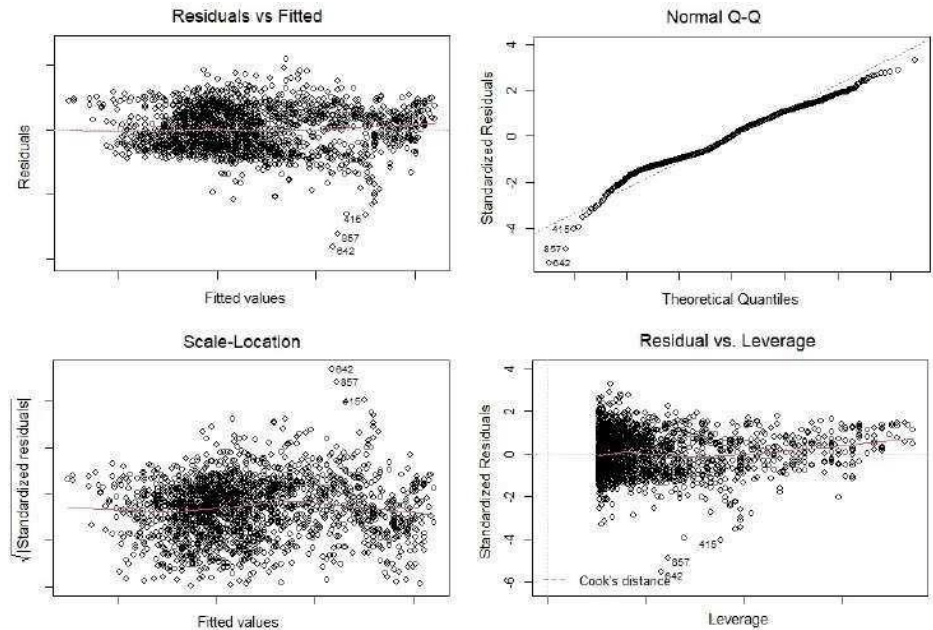

Figure 13. The diagnostic plots of time series linear regression model for February Land Surface Temperature (LST) data in figure (4).

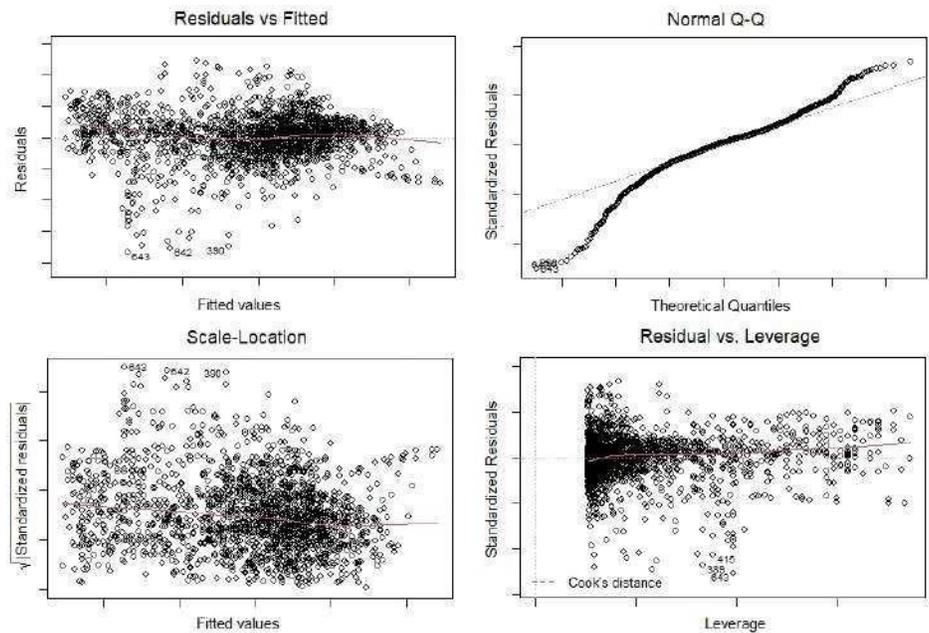

Figure 14. The diagnostic plots of time series linear regression model for April Land Surface Temperature (LST) data in figure (4).

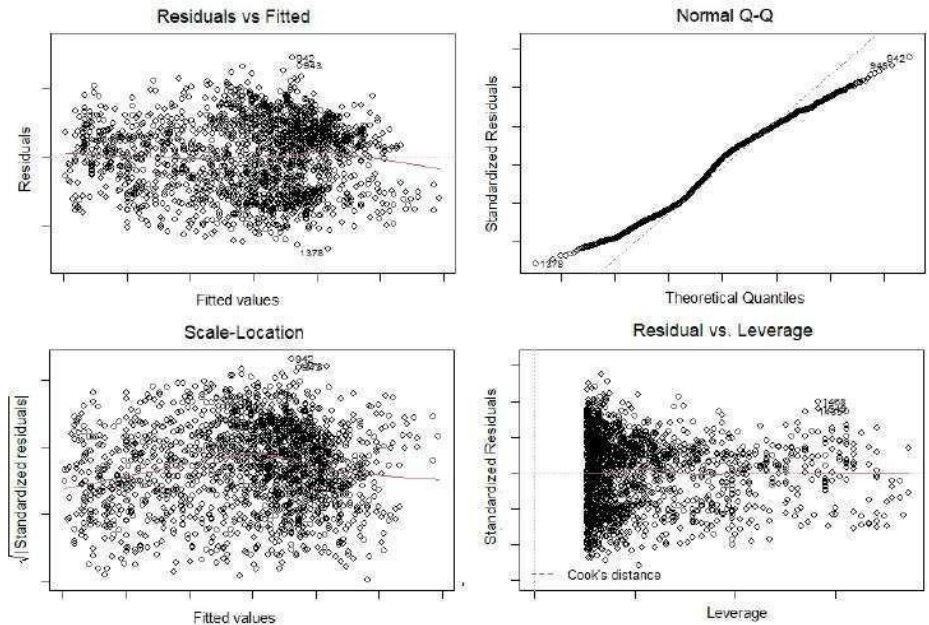

Figure 15. The diagnostic plots of time series linear regression model for July Land Surface Temperature (LST) data in figure (4).

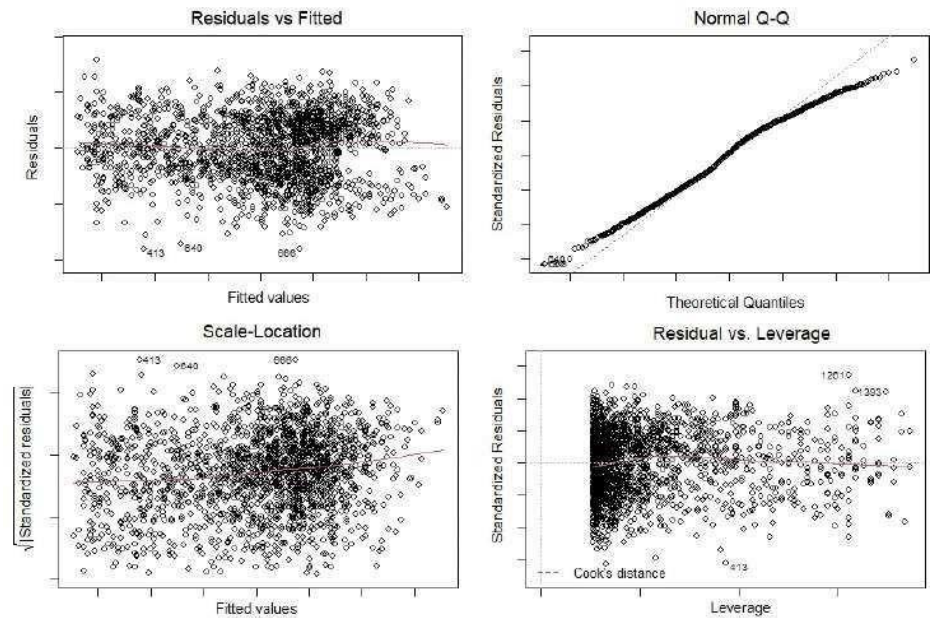

Figure 16. The diagnostic plots of time series linear regression model for October Land Surface Temperature (LST) data in figure (4).

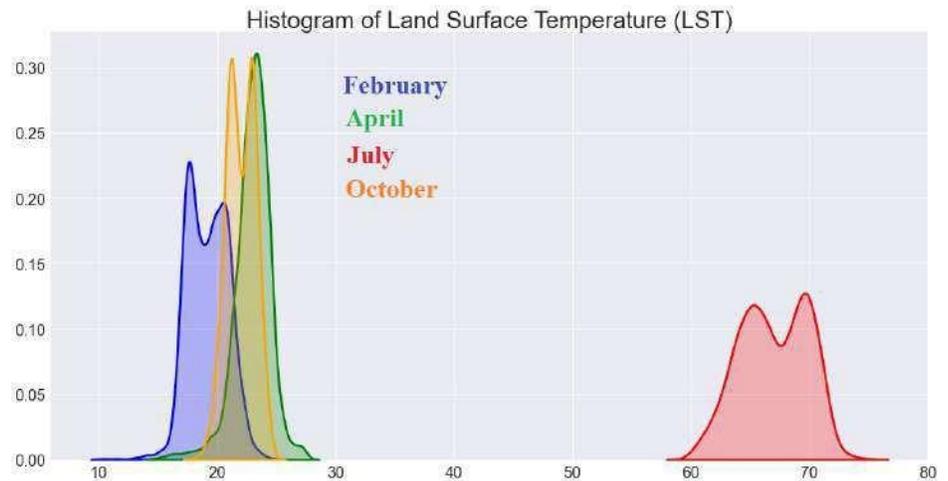

Figure 17. Histogram of (LST) values during February, April, July and October (2021) in figure (4).